\documentclass{ws-p8-50x6-00}
\begin{document}

\title{Baryon Transport in Dual Models and 
  the Possibility of a Backward Peak in Diffraction }

\author{Fritz W. Bopp} 

\address{Universit{\"a}t Siegen, Fachbereich Physik,  D--57068 Siegen, Germany
\\ bopp@physik.uni-siegen.de }


\maketitle

\vspace*{-4.2cm}
{\hfill {SI-01-2}}
\vspace*{+3.6cm}

\abstracts{
Dual string models contain significant baryon transfers and seem essentially
consistent with the available data. We here turn to a careful consideration
of the relevant topological structures. The baryon transfer is associated with
one of two possible types of cuts in various baryonium exchanges. As the baryonium
with the highest intercept easily couples to two Pomerons such transfers should
occur abundantly in percolating dense Pomerons systems. From the color structure
this quark-less baryonium can be identified with an Odderon exchange.
As the Odderon is predicted to have an almost Pomeron-like trajectory
it has to involve small coupling constants so that steeper trajectories can
initially determine the data. As this suppression is not anticipated for diffractive
processes a tiny observable backward peak should occur in the initial baryon
distribution for massive diffractive systems. }
\noindent {\bf  Baryon transfer in particle scattering}

The suppression of the long range transfer of baryon charges in inclusive spectra
and in annihilation is in the range below \( \sqrt{\textrm{s}}=10 \) GeV resp.
\( \Delta y=2 \) determined by a baryonium intercept of 
\( \alpha _{Transfer}-\alpha _{Pomeron}=-1 \) \cite{1,2,3}. At the center of ISR 
\cite{4,5} there is an indication of a
flattening. New preliminary data from the H1 experiment at HERA \cite{6} support
this turnover. The trajectories required by a HERA ratio compared with its
ISR value can be estimated as (see also \cite{7}) \vspace{-0.3cm}
\[
\alpha _{Transfer}-\alpha _{Pomeron}=-0.4\pm 0.2\: .\]
Both obtained slopes correspond to the classical Dual Topological model \cite{8}
expectation\cite{9} \( \alpha _{junction}^{I}-1=-1.0 \) and \( \alpha _{junction}^{0}-1=-0.5 \)
for both trajectories. However the value of the final trajectory is rather uncertain.
Values of \( \alpha _{junction}^{0}-1=-0.8\: ...\: 0 \) can be found in the
literature \cite{10,11,12}. 

Such baryonium trajectories are included in most fragmentation codes in a somewhat
indirect way (see e.g. \cite{13,14,15}). The splitting functions usually contain
all possible quark and diquark transition. It includes a pure diquark contribution
which corresponds to baryonium cut. In the widely used JETSET code \cite{16}
the combinatoric suppression is tuned to yield effectively the initial steep
slope.

\noindent {\bf Concepts for slowing-down initial baryons in heavy ion scattering}

To understand the data it seems necessary to include interplay of string if
they get sufficiently dense in transverse space. It was proposed that there
are new special strings \cite{20,21}. In contrast, we shall here maintain the
general factorization hypothesis between initial scattering in the quark phase
and the final hadronization within standard strings. Final state
interactions
are known to introduce some correction to the simple picture.

An obvious mechanism involves the incoming baryons. The usual Pomeron exchange
in the Dual Parton model leaves a quark and a diquark for the string ends. Diquarks
are no fixed entities and multiple scattering processes can split them in a
conventional two Pomeron interaction \cite{22,7,23,24,25}. It is natural to
expect that diquark break-ups considerably slow down the baryons evolving. The
probability for such an essentially un-absorbed \cite{26} process strongly
depends on the density\cite{27,28,29,22}. As required by the experiment this
is a drastic effect for heavy ion scattering \cite{22} while for hadron-hadron
scattering multiple scattering is sufficiently rare to preserve the known hadron-hadron
phenomenology \cite{25}. 

The behavior of the baryon quantum number slowed down by such a break-up is
not trivial. In topological models the baryon contains Y-shaped color electric
fluxes connected by a vortex line. The energy distribution of quarks with vortex
lines (or of the fully separated vortex lines) in the structure function is
a priory not known and requires special consideration.

\noindent {\bf Special baryon transfers in the Topological model}

For this question we return to the Dual Topological model \cite{9} on
which models like the Dual Parton model are based  
(for a recent discussion on baryonium see also  \cite{30}) and 
emphasize topological aspects. 
The Pomeron exchange corresponds to a cylinder
connecting the two scattering hadrons. On an arbitrary plane intersecting this
t-channel exchange the intersection is topologically a circle. More specifically
amplitudes with clockwise respectively anticlockwise orientation have to be
considered and the cylinders or the circles come with two orientations.
This distinction is usually not very important as it is always topologically
possible to attach hadrons in a matching way. Except for \( C \)-parity conservation
(which follows from cancellations) no special restrictions result. 

Pomerons have a transverse extent and if they are close in transverse space
they should interact. Hadronic interaction is sufficiently strong to be largely
determined by geometry as long as there is no mechanism of suppression. It is
therefore reasonable to expect that the coupling does not strongly depend on
orientation. 

The two distinct configurations occur. Two Pomerons with the same orientation
can if they touch (starting locally at one point in the exchange-channel time)
shorten their circumference and form a single circle:
\vspace*{0.2cm}
{\par\centering \resizebox*{0.5\textwidth}{0.03\textheight}{\includegraphics{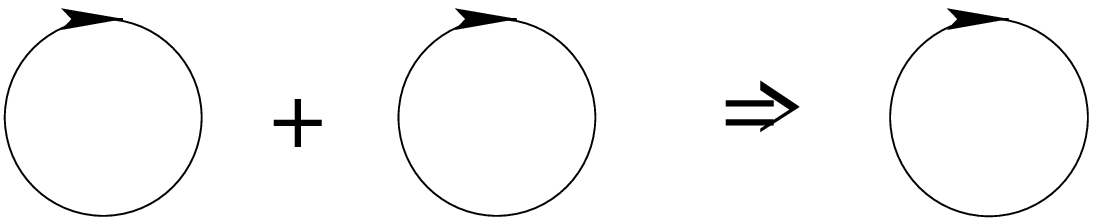}} \par}

\noindent This then corresponds to the usual triple Pomeron coupling experimentally
well-known from diffractive processes. 

For two Pomerons with opposite orientation the situation is more complicated.
Like for soap bubbles the two surfaces which get in contact can merge and form
a single membrane, starting locally with the creation of a vortex pair. The
joining inverts the orientation of the membrane. On the intersecting plane one
now obtains -- instead of the single circle -- three lines originating in a
vortex point and ending in an anti-vortex point as shown below: 
\vspace*{0.2cm}
{\par\centering \resizebox*{0.5\textwidth}{0.03\textheight}{\includegraphics{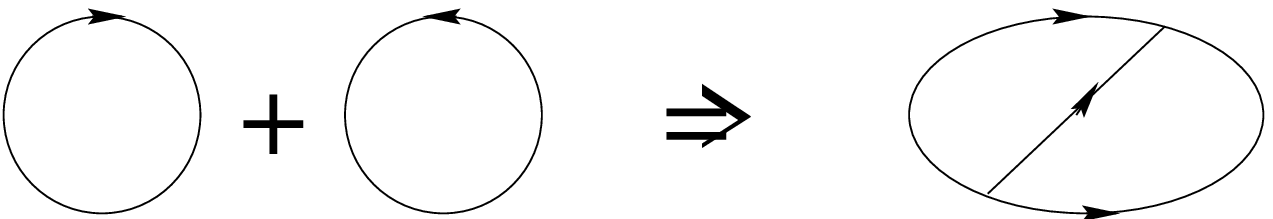}} \par}

\noindent Lacking a topological name for the object the term membraned cylinder
will be used in the following.

How do this membraned cylinder contribute to particle production? Similar to
the triple Pomeron case there are three different ways to cut through a membraned
cylinder:
\vspace*{0.0cm}
{\par\noindent \centering \resizebox*{0.22\textwidth}{0.05\textheight}{\includegraphics{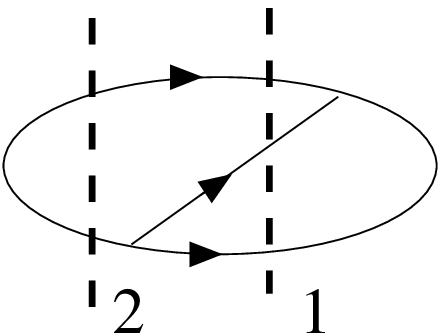}} \par}

\noindent The symmetric cut (numbered 1) which also intersects the membrane has
vortex lines on both sides. They present a topological description of the baryon
transfers considered above. By symmetry they contribute with a positive sign.
Cuts which intersect only two sheets (numbered 2) contribute to the two string
contributions. Their sign is unknown. As they contain a closed internal fermion
(vortex line) loop we here assume a negative sign.

\noindent {\bf The identification with the Odderon}

It is widely believed that calculable hard processes can be used as a guide
to model corresponding soft processes as a suitable extrapolation. 

The topological considerations in perturbative QCD are based on the \( 1/N_{C} \)
- expansion. This approximation selects contributions according to the magnitude
of their color factors reflecting "coloring" choices of suitably drawn color lines.
For an amplitude of a given structure with a given number of couplings the leading
order \( 1/N_{C} \) contribution can be drawn without crossing color lines.
In special situations the drawing has to be done on topological structures which
are more complicated than the simple plane considered above. An example is the
cylinder which is assumed to be responsible for the Pomeron contribution. 

The known example of the soft hard correspondence is the connection between
soft and hard Pomerons. To identify the hard partner of the soft Pomeron we
first observe that the simplest representation of a Pomeron in PQCD involves
the exchange of two gluons which form a color singlet with the required positive
charge parity. Following this concept it can be shown \cite{31} that a generalization
of such an exchange gives the dominant contribution at very high energies in
a well defined approximation. It is called ``hard'' or BFKL Pomeron and involves
a ladder of two exchanged Reggeized gluons linked by a number of gluons. In
the topological expansion the leading structure of a BFKL Pomeron corresponds
to a cylinder. The two basic Reggeized gluons are exchanged on opposite sides
parallel to the axis: 
\vspace*{0.2cm}
{\par\centering \resizebox*{0.33\textwidth}{0.05\textheight}{\includegraphics{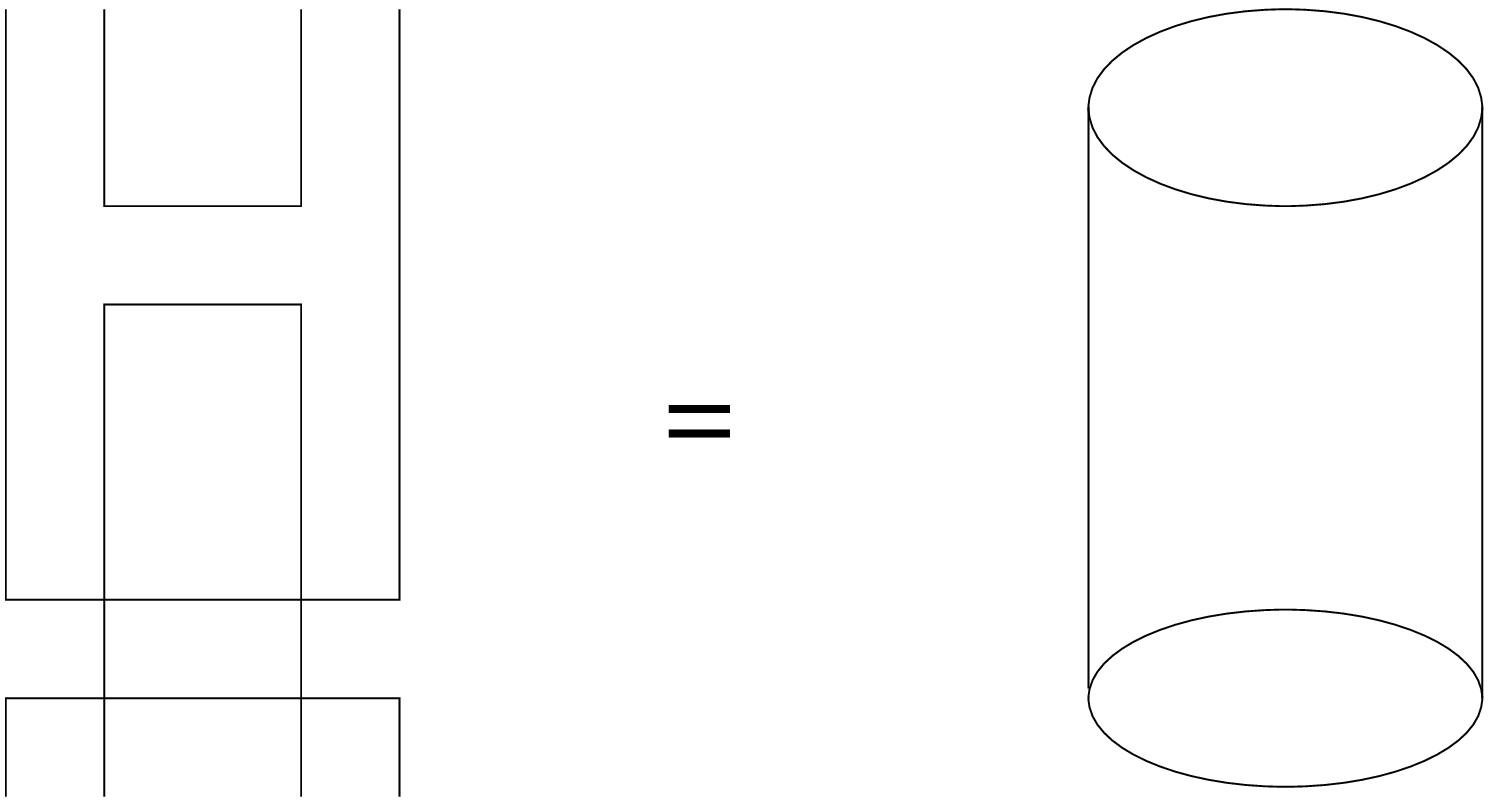}} \par}

\noindent Their matching inner color lines can be linked in front of the cylinder
without color line crossing. Analogously their matching outer lines can be connected
on the back of the cylinder. 

Going back to the soft regime the basic assumption
in topological models is that the \( 1/N \)-expansion stays valid and that
the soft Pomeron therefore maintains its cylindrical structure needed for the
two string phenomenology of hadronic final states. If cut, soft and hard Pomerons
lead to similar two string final states. As difference it remains
that the trajectory of the observed soft Pomeron is just shifted downward roughly
by a third of a unit from hard Pomeron calculated in leading logarithmic approximation.

Can one find a similar connection for the membraned cylinder? The simplest representation
spanning such a topological structure involves three gluons, one on each sheet
exchanged parallel to the axis. Any gluon linking these exchanges has then to
pass through a vortex line in which the three sheets join. In the \( 1/N \)
expansion extended to baryons this means that the color lines have to cross
passing this line. The basic QCD structure of the membraned cylinder exchange
is therefore the following:
\vspace*{0.2cm}
{\par\centering \resizebox*{0.33\textwidth}{2cm}{\includegraphics{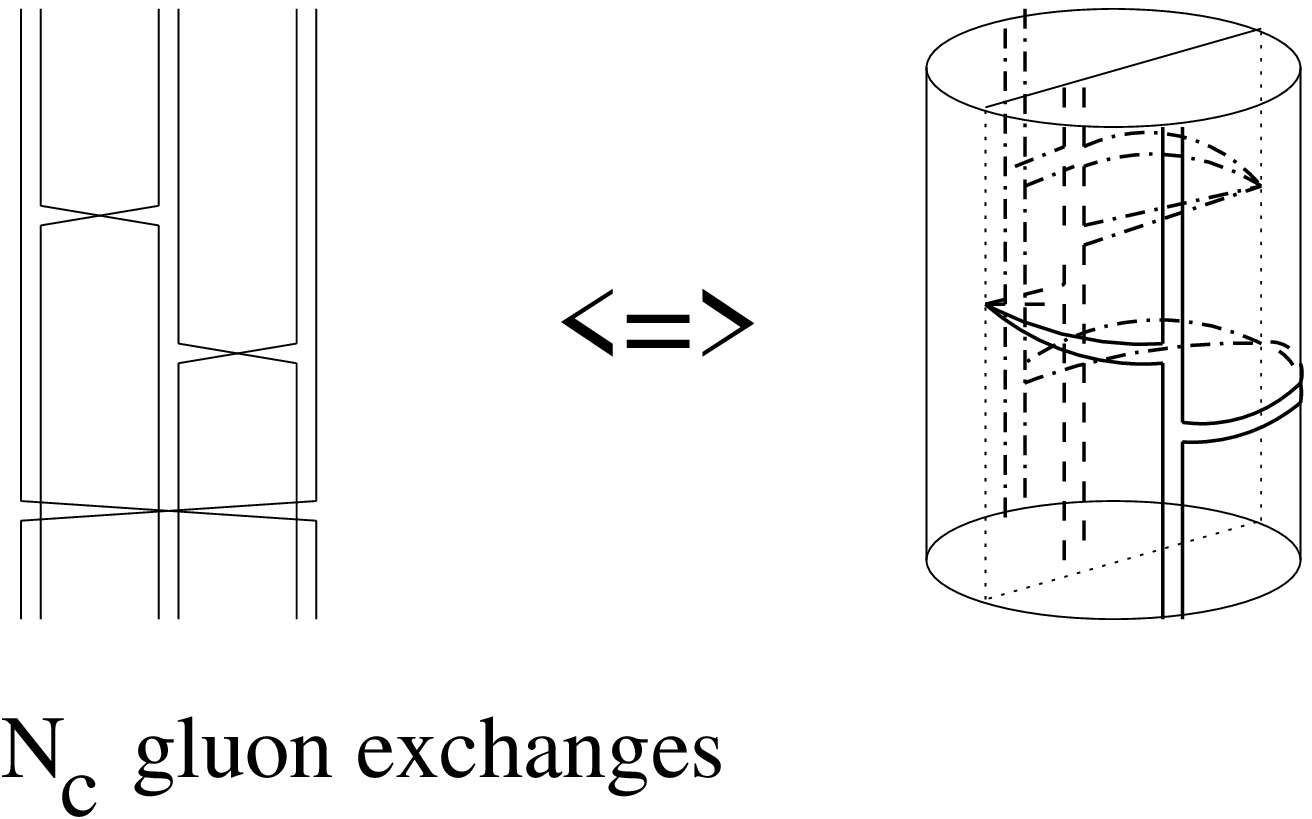}} \par}
\vspace{0.2cm}

Looking from the other side a color singlet of three gluons can have the quantum
numbers of a Pomeron or an Odderon \cite{32}. There is a simple topological
property of the Odderon. A single uncrossed gluon link would project the color
structure of the pair of exchanged Reggeized gluons to that of a single gluon,
\( (8)_{F} \), and the exchange would have to correspond to a Pomeron-like
contribution. The Odderon will therefore have to involve crossed links. Hence
it has exactly the topology of the membraned cylinder.

To visualize the baryonic color structure of the Odderon with its crossed exchanges
one can replace the exchanged gluons by quark antiquark pairs without changing
color lines. The so modified membraned cylinder just represents an exchanged
baryon antibaryon pair.

In the QCD approximation used for the ``hard'' Pomeron the properties of the
``hard'' or BKP Odderon \cite{33} were calculated and the predicted intercept
is \( 0.96 \) \cite{34} or \( 1.0 \) \cite{lipatov99}, depending of the
details of the considered state. Again a mismatch between this hard leading
logarithmic Odderon intercept and the possibly experimentally observed soft
value by about a third is indicated by the data.

\noindent {\bf General  consequences  of membraned cylinder exchanges}

Our preferred hypothesis is that the membraned cylinder exchange has a small
or almost vanishing imaginary part. In this way there are no constraints from
total cross section fits. Also, there is no coupling of the
total C odd Odderon on a C even Pomeron pair. The cancellation allows a small
or vanishing Odderon to contain sizable individual components of opposite
sign, which (by looking at baryon exchange) can be used to determine the
soft Oddderon trajectory.

In heavy ion scattering where the Pomerons are dense in transverse space they
can join and form a Pomeron or an Odderon. The individual strings are no longer
independent but the general picture of particle production in separate universal
strings survives. The probability of such an interaction of strings is growing
proportional to the density and eventually to rapidity range. The transition
from a Pomeron pair to the centrally cut membraned cylinder involves baryon
antibaryon pair production, which should occur quite abundantly. Between a proton
and  Pomerons the cut membraned-cylinder is a very efficient mechanism of baryon
stopping. 

Both effects correspond to experimental observations. As the trajectory is not
well determined it is hard to obtain really reliable quantitative estimates
which can be tested convincingly with on heavy ion data.

\noindent {\bf The backward peak in diffraction and possibly in electro production}

There is however a very specific qualitative prediction which can be tested.
Consider a massive diffractive system. Usually the diffractively produced particles
will originate in two strings of a cut Pomeron and the baryon charge will stay
on the side of the initial proton. As usual there might be some migration to
the center with a slope in rapidity eventually corresponding to the difference
of the Odderon and the Pomeron trajectory. Topologically it involves a horizontal
cut through the following structure:
\vspace{0.2cm}
{\par\centering \resizebox*{!}{0.08\textheight}{\includegraphics{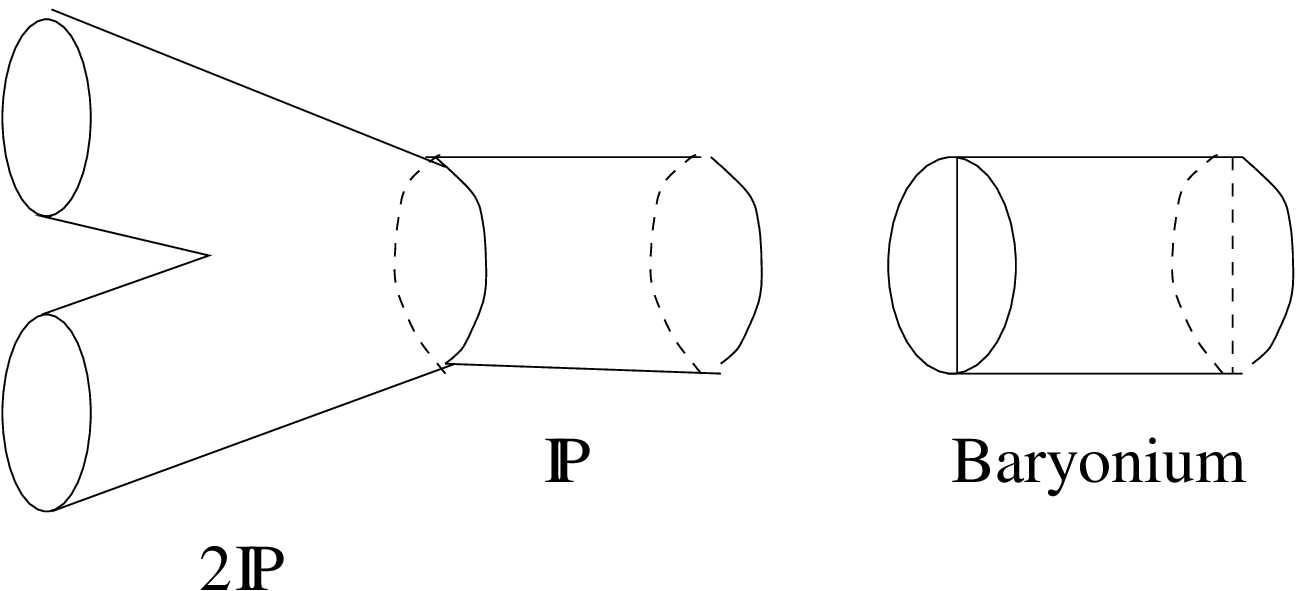}} \par}
\vspace{0.1cm}
The high Odderon trajectory requires a clear suppression from the coupling constants
to stay consistent with low energy data. The natural candidate for such a suppression
is the Pomeron-Baryonium vertex, which involves no large overlap and for which
cancellation between separate contributions can be expected. In consequence
at a certain distance it should be more favorable for the membraned cylinder
to span the total diffractive region and to utilize the more favorable coupling
to the two Pomerons. In this way the initial baryon will end up exactly at the
backward end of the diffractive system. 
\vspace{0.2cm}
{\par\centering \resizebox*{!}{0.19\textheight}{\includegraphics{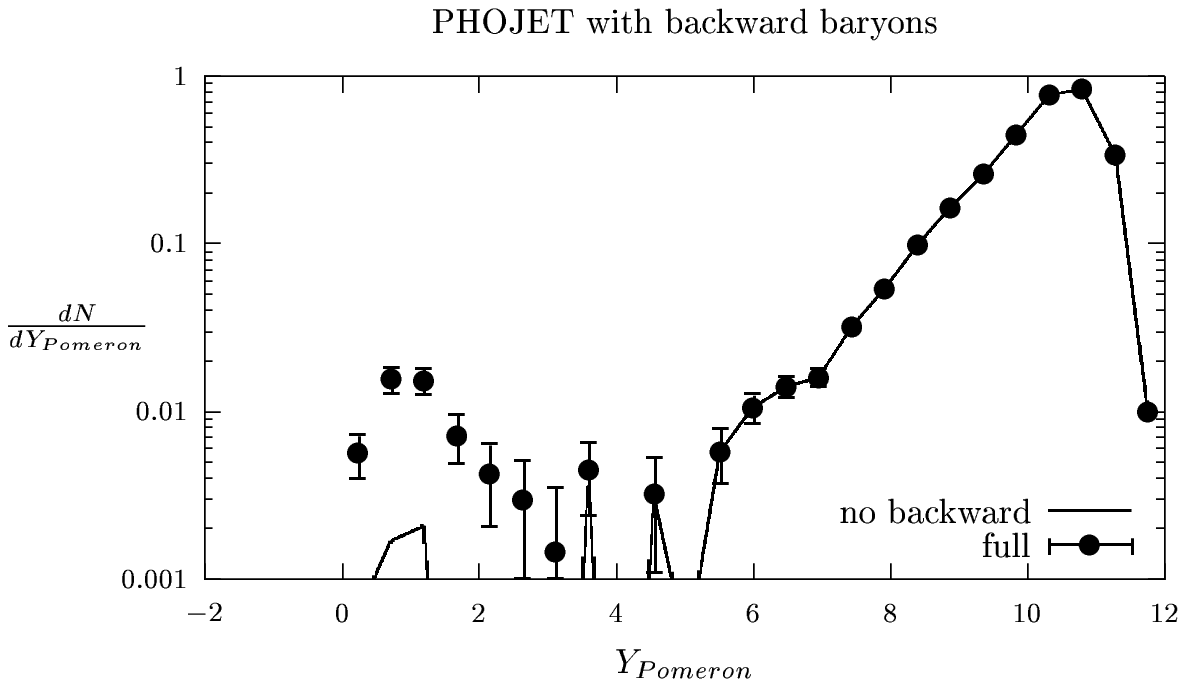}} \par}
\vspace{0.1cm}
It should be visible if one plots the rapidity distribution in relation to the
inner end of the diffractive region, i.e. as function of 
\( y_{\{Pomeron\}}=y_{\{CMS\}}-\ln (m\sqrt{s}/M(diffr.)^{2}) \).
To illustrate the expected small backward peak we show the result of a calculation
with the PHOJET Monte Carlo code \cite{35} of the incoming proton spectrum
for diffractive events with a mass of \( 300 \) GeV for \( pp \)-scattering
of \( 1.8 \) TeV with standard parameters below. To select diffractive events
a lower cutoff of \( x_{F}=0.95 \) was used.
PHOJET contains diquark exchanges and yields reasonable baryon spectra in the
forward region. To obtain the postulated backward peak we just mixed in a suitable
sample of inverted events (with disabled diquark exchanges\cite{36} ).

\end{document}